\documentclass[journal]{IEEEtran}
\usepackage{multirow}
\usepackage{graphicx}          % Include this line if your 
\usepackage[table]{xcolor}
\usepackage{float}
\usepackage{bm}
\usepackage{amsmath}
\usepackage{amsfonts}
\usepackage{subfigure}

\numberwithin{lem}{section}

\numberwithin{cor}{section}

\newtheorem{thm}{Theorem}
\numberwithin{thm}{section}

\newtheorem{rem}{Remark}
\numberwithin{rem}{section}

\numberwithin{defn}{section}

\usepackage{algorithm, algorithmicx, algpseudocode}
\usepackage{color}
\usepackage{cite}
\allowdisplaybreaks

\graphicspath{{./Figures/Irrigation/}{./Figures/SMD/}}

\begin{document}

\title{Structured preconditioning of conjugate gradients for
  path-graph network optimal control problems\thanks{Supported in part by the
    Australian Research Council (LP160100666).}}
%  Optimal control of path-graph networks: Decomposable conjugate gradient preconditioning for second-order methods}

\author{Armaghan~Zafar, Michael~Cantoni,
  and~Farhad~Farokhi% <-this % stops a space
  \thanks{AZ, MC, and FF are with the Department of Electrical and
    Electronic Engineering, The University of Melbourne, Parkville VIC
    3010, Australia. (emails:
    \{armaghanz@student;~cantoni@;~farhad.farokhi@\}.unimelb.edu.au)}
}%

	\maketitle

	\begin{abstract} 
		  A structured preconditioned conjugate gradient (PCG) solver
          is developed for the Newton steps in second-order methods
          for a class of constrained network optimal control
          problems. Of specific interest are problems with
          discrete-time dynamics arising from the path-graph
          interconnection of $N$ heterogeneous sub-systems. The
          computational complexity of each PGC step is shown to be
          $O(NT)$, where $T$ is the length of the time horizon. The
          proposed preconditioning involves a fixed number of block
          Jacobi iterations per PCG step. A decreasing analytic bound on
          the effective conditioning is given in terms of this number.
          The computations are decomposable across the
          spatial and temporal dimensions of the optimal control
          problem,
          into sub-problems of size independent of $N$
          and $T$. Numerical results are provided for
          a mass-spring-damper chain.
	\end{abstract}

	\begin{IEEEkeywords}
          Optimal control of
          networks; Structured second-order solver; System chains.
%		Enter key words or phrases in alphabetical 
%		order, separated by commas. For a list of suggested keywords, send a blank 
%		e-mail to keywords@ieee.org or visit \underline
%		{http://www.ieee.org/organizations/pubs/ani\_prod/keywrd98.txt}
	\end{IEEEkeywords}
	
	\section{Introduction}\label{sec:introduction}
        \IEEEPARstart{C}{onsider} the path-graph interconnection of
        $N$ heterogeneous sub-systems with dynamics given by
        \begin{equation}\label{eq:discrete_time_dynamics_cascaded_system}
          x_{j,t+1} = A_{j,t}x_{j,t} + B_{j,t}u_{j,t} + E_{j,t}x_{j-1,t} + F_{j,t}x_{j+1,t},
        \end{equation}
        where $x_{j,t}\in \mathbb{R}^{n_{j}}$ and
        $u_{j,t}\in \mathbb{R}^{m_{j}}$ are the state and input of
        sub-system $j\in\mathcal{N}= \{1,2,...,N\}$ at time
        $t \in \mathcal{T} = \{0,1,...,T\}$,
        respectively. The initial conditions are given by
        $x_{j,0} = \boldsymbol{\xi}_j \in \mathbb{R}^{n_j}$ for $j\in\mathcal{N}$
        and the spatial boundary conditions are given by
        $x_{0,t}=\boldsymbol{\chi}_t \in\mathbb{R}^{n_0}$ and
        $x_{N\!+\!1,t} = \boldsymbol{\zeta}_t \in\mathbb{R}^{n_{N\!+\!1}}$
        for $t\in\mathcal{T}$.
        The constrained finite-horizon
        linear-quadratic (LQ) optimal control problem of interest is
        the following: {
          \begin{subequations}\label{eq:LQ_optimal_control_problem}
            \begin{equation}\label{eq:cost_function}
              %\begin{split}
                \min_{\substack{(x_{j,t})_{(j,t)\in(\!\{0,N\!+\!1\}\cup\mathcal{N})
                      \!\times\!\mathcal{T}}\\
                    (u_{j,t})_{(j,t)\in\mathcal{N}\!\times\!%(
                      \mathcal{T}}
                      %\backslash\{T\})}
                  }} \frac{1}{2}
                \sum_{j\in\mathcal{N}}
                %\Biggl(
                \sum_{t\in\mathcal{T}}
                  %\mathcal{T}\backslash
                  %T}
                \ell_{j,t}(            
                  x_{j,t},
                  u_{j,t})
            	\end{equation}
            \text{subject to}
            \begin{align}
              % x_{j,t+1} &= A_{j,t}x_{j,t} + B_{j,t}u_{j,t} +
              % E_{j,t}x_{j-1,t} + F_{j,t}x_{j+1,t},\notag\\
                          \eqref{eq:discrete_time_dynamics_cascaded_system} %\qquad
                          &~\text{ for }
                            (j,t)\in\mathcal{N}\times(\mathcal{T}\backslash\{T\}),
                            \label{eq:equality_constraint}\\
                          x_{0,t} =\boldsymbol{\chi}_t,\ x_{N+1,t}=\boldsymbol{\zeta}_t %\qquad
                            &~\text{ for } t \in
                                                 \mathcal{T},
                                                \label{eq:boundary_condition}\\	
                          x_{j,0} = \boldsymbol{\xi}_j %\qquad
                            &~\text{ for } j \in
              \mathcal{N}, \label{eq:initial_condition}\\ 
                          C_{j,t}x_{j,t} + D_{j,t}u_{j,t} \leq \boldsymbol{\kappa}_{j,t} %\qquad
                            &~\text{ for }
                                                (j,t)\in\mathcal{N}\times
              \mathcal{T}, 
                             \label{eq:inequality_constraint}
            \end{align}
          \end{subequations}
        }%
        where $\ell_{j,t}(x,u) = x^\prime Q_{j,t} x + 2 x^\prime
        S_{j,t} u + u^\prime R_{j,t} u$, 
        $C_{j,t} \in \mathbb{R}^{\nu_j \times n_j}$,
        $D_{j,t} \in \mathbb{R}^{\nu_j \times m_j}$ and
        $\boldsymbol{\kappa}_{j,t} \in \mathbb{R}^{\nu_j}$. For $j\in\mathcal{N}$ and
        $t\in\mathcal{T}\backslash\{T\}$, it is assumed that
        $ Q_{j,t} = Q_{j,t}^{\prime} \succeq 0$,
        %$\in \mathbb{R}^{n_j
        %  \times n_j}$,
        $R_{j,t} = R_{j,t}^{\prime}\succ 0$,
        %$\in \mathbb{R}^{m_j \times
        % m_j}$,
        and
        % \begin{equation}\label{eq:positive_semidefinite_cost}
        %   \begin{bmatrix}
        %     Q_{j,t} & S_{j,t}'\\
        %     S_{j,t} & R_{j,t}
        %   \end{bmatrix} \succeq 0,
        % \end{equation}
        %so
        %that 
 %       which along with
 %       \eqref{eq:positive_semidefinite_cost} implies that
        $Q_{j,t} - S_{j,t}^{\prime}R_{j,t}^{-1}S_{j,t} \succeq
        0$. Moreover, for every $j\in\mathcal{N}$, $Q_{j,T}\succeq 0$,
        but $S_{j,T}=0$, $R_{j,T}=0$, and $D_{j,T}=0$, so that
        $u_{j,T}$ plays no role (i.e., it can be removed as a decision
        variable.) Under these assumptions the problem
        \eqref{eq:LQ_optimal_control_problem} is a convex quadratic
        program with $O(NT)$ decision variables and $O(NT)$ constraints.

        While the cost \eqref{eq:cost_function} and inequality
        constraints \eqref{eq:inequality_constraint} are separable
        across the sub-systems and time horizon, there is coupling in
        the equality constraint
        \eqref{eq:equality_constraint}. Specifically, there is
        spatial coupling between states of adjacent sub-systems, and
        inter-temporal coupling.
        %i.e., $j,~j-1,~j+1$ and adjacent time
        %indices $t,\ t+1$ in dynamics of the cascaded system
        %\eqref{eq:equality_constraint}. Such
        Path-graph network dynamics of this kind are
        relevant in the operation of irrigation
        channels~\cite{Nasir2019stochastic},
        %~\cite{cantoni2007control},
        %~\cite{li2005water},
        vehicle platoons~\cite{zheng2016distributed}, supply chains
        \cite{perea2003model}, and radial power networks
        \cite{Giannitrapani2017optimal}. The
        structure also arises from the discretization of one-dimensional
        partial differential equations~\cite{Rees2010optimal}.
        %in optimization problems subject to
        %discretized one dimensional elliptic partial differential
        %equations; see
        %\cite{Maurer2000a,Rees2010optimal}.

        This note is about the computation of second-order search
        directions for solving the quadratic program
        \eqref{eq:LQ_optimal_control_problem}. Specifically, a
        preconditioned conjugate gradient (PCG) solver (e.g.,
        see~\cite{Hackbusch2016iterative}) is developed for the
        Newton steps in second-order methods, such as the interior
        point method~\cite{Nocedal2000numerical}. The main innovation
        pertains to the $O(NT)$ computational complexity of each PCG
        iteration, and decomposability of the preconditioning
        computations across both the temporal and spatial dimensions,
        into sub-problems of sizes that are independent of $N$ and
        $T$. The computations are amenable to implementation as
        $\lceil N/2\rceil$ parallel threads each comprising a sequence
        of $2T$ (possibly dense but small) sub-problems.

        Structure in second-order methods for optimal
        control problems was studied
        in~\cite{Wright1993interior,Rao1998application}, where the
        so-called Riccati-factorization approach was originally
        developed, and more recently in~\cite{wills2003interior,
          Wang2010fast, Shahzad2012stable, domahidi2012efficient,
          Nielsen2019direct}. These papers all focus on the structure
        associated with localized coupling in the temporal dimension
        of optimal control problems. Following the underlying approach
        for problem (\ref{eq:LQ_optimal_control_problem}) results in
        solvers with $O(TN^3)$ computational complexity for each of
        the moderate number of Newton steps needed for second-order methods
        to converge (typically $10-20$ steps). The computations
        are decomposable across the temporal dimension, but not the
        spatial dimension. The resulting sub-problems, of size $O(N)$,
        are amenable to distribution across parallel processors in a
        tree type communication network, leading to $O(\log(T)N^3)$ time
        complexity~\cite{Nielsen2019direct}.
        
        In~\cite{Cantoni2017structured}, the aforementioned approach
        is pursued in the special case of
        (\ref{eq:LQ_optimal_control_problem}) with directed spatial
        coupling, by interchanging the role of the time and space
        indexes to develop a Newton step solver with computational
        complexity $O(NT^3)$. The computations are decomposable across
        the spatial dimension of the problem, but not the temporal
        dimension. Again, parallel processing can lead to
        $O(\log(N)T^3)$ time complexity.

        All of the approaches described above constitute direct
        methods for solving the Newton steps. In particular, all are
        related, in some way, to structured block-LU factorization for
        a permutation of variables that yields a block tri-diagonal
        structure in the linear system of equations to be
        solved. With direct methods, it appears to be difficult to
        leverage both the spatial and the temporal structure
        in (\ref{eq:LQ_optimal_control_problem}).

        The proposed PCG method is an iterative solver, of the kind
        used for large sparse
        problems~\cite{Hackbusch2016iterative}. For
        (\ref{eq:LQ_optimal_control_problem}), the size of the linear
        equation to solve at each Newton step is $O(NT)$. Thus, in the worst
        case, it may take $O(NT)$ iterations to terminate. It is
        well-known that preconditioning can significantly reduce the
        number of PCG iterations needed. In this note, it is
        proposed to use a fixed number of block Jacobi iterations for
        preconditioning. In principle, this fixed number can be
        selected to achieve preconditioning specifications, in that a decreasing
        analytic bound on the conditioning of the outcome is
        provided. For the numerical example presented, it is observed
        that as few as two Jacobi iterations can result in a much
        smaller number of PCG steps than the worst-case bound
        described above. Importantly, the preconditioning steps are
        decomposable across both the spatial and temporal
        dimension of \eqref{eq:LQ_optimal_control_problem}. The size
        of the resulting $O(NT)$ parallelizable sub-problems is independent
        of $N$ and $T$. As such, the computational complexity of PCG
        steps is $O(NT)$. In the worst-case of $O(NT)$ iterations,
        the computational complexity of a Newton step becomes
        $O(N^2T^2)$. So for $T \approx N$, as perhaps required for
        the optimal control problem to be meaningful, the proposed
        approach is (at the least) no worse than the structured direct
        methods discussed, and potentially much better for large
        problems.

        First-order methods can also lead to
        structured solvers for separable-in-cost quadratic programs
        like \eqref{eq:LQ_optimal_control_problem}. For example,
        methods based on dual
        decomposition~\cite{Bertsekas1997parallel}, and operator
        splitting methods such as ADMM~\cite{Boyd2011distributed} and
        FAMA~\cite{stathopoulos2016operator} can lead to simple
        parallelizable computations. For the structure in
        \eqref{eq:LQ_optimal_control_problem}, the dual decomposition
        technique of \cite{Falsone2017dual} leads to local
        computations for each sub-system. Similarly, the ADMM
        approach presented in \cite{Cantoni2018scalable}, and projected
        sub-gradient algorithm of \cite{Nedic2010constrained}, also
        yield decomposable computations. However, these first-order
        methods typically require a huge number of iterations to
        converge. The issue is exacerbated within the path-graph
        context of this note, since the algebraic connectivity of the
        underlying sparsity pattern, which influences the rate of
        convergence~\cite{Franca2017how,Nedic2018network}, tends to
        zero as $N$ grows. This
        motivates the consideration of second-order methods. The
        challenge is to maintain structure in the computations.

		The note is organized as follows. An equivalent re-formulation
        of problem \eqref{eq:LQ_optimal_control_problem} is presented in
        Section~\ref{sec:problem_formulation}, including the structure
        of corresponding Newton steps in
        Section~\ref{subsec:newton_method}. PCG methods are overviewed
        in Section~\ref{sec:pcg_method}, and the structured
        preconditioner based on fixed block Jacobi iterations is
        developed in
        Section~\ref{sec:structured_preconditioner_for_CG_method}. The
        proposed PCG algorithm is explored numerically for
        mass-spring-damper chain example in
        Section~\ref{sec:numerical_results}. Concluding remarks are
        provided in Section~\ref{sec:conclulsions}.
	%end of section
	\section*{Notation}\label{sec:notations}
        Identity matrices are denoted by
        $I$. $\mathrm{blkdiag}(\cdot)$ denotes the matrix with block
        diagonal elements given by the arguments, which are the only
        non-zero elements, and $\mathrm{col}(\cdot)$ denotes the
        concatenation of the input arguments into a column vector.
        Every block tri-diagonal matrix is parameterized by sequences
        $\Phi=(\Phi_k)_{k=1}^m \in \prod_{k=1}^m\mathbb{R}^{l_k\times
          l_k}$ and
        $\Omega=(\Omega_k)_{k=2}^m \in
        \prod_{k=2}^{m}\mathbb{R}^{l_{k-1}\times l_k}$ for
        appropriate $(l_k)_{k=1}^{m}\subset\mathbb{N}^{m}$ and
        $m\in\mathbb{N}$. Given such sequences $\Phi$ and $\Omega$,
        the corresponding block tri-diagonal matrix is denoted by
		\begin{equation*}
			\mathrm{blktrid}(\Phi,\Omega) = 
			\left[
				\begin{array}{cccc}
					\Phi_1 &\Omega_2^{\prime} & &\\
					\Omega_2 &\Phi_2 &\ddots &\\
					&\ddots &\ddots &\Omega_{m}^{\prime}\\
					& &\Omega_{m} &\Phi_{m}\\
				\end{array}
			\right] \in \mathbb{R}^{\bar{l}\times \bar{l}},
                      \end{equation*}
                      where $\bar{l}=\sum_{k=1}^m l_k$.
	%end of section
	\section{Problem Re-Formulation}\label{sec:problem_formulation}
        Defining
        ${u}_j = \mathrm{col}(u_{j,0}, \ldots, u_{j,T-1}) \in
        \mathbb{R}^{m_j T}$,
        ${x}_j = \mathrm{col}(x_{j,0}, \ldots, x_{j,T}) \in
        \mathbb{R}^{n_j(T+1)}$, and slack variables
        ${\theta}_j = \mathrm{col}(\theta_{j,0}, \ldots,
        \theta_{j,T}) \in \mathbb{R}^{\nu_j(T+1)}$,
        problem \eqref{eq:LQ_optimal_control_problem} can be
        reformulated as the following quadratic program:
		\begin{subequations}\label{eq:temporal_stacked_problem}
			\begin{equation}\label{eq:stacked_cost}
				\min_{\substack{(x_j)_{j\in\{0,{N\!+\!1}\}\cup\mathcal{N}}\\
								(u_j)_{j\in\mathcal{N}}}}
                                                            \frac{1}{2}
                                                            \sum_{j\in\mathcal{N}}
				\begin{bmatrix}
				x_{j}\\
				u_{j}
				\end{bmatrix}^{\prime}
				\begin{bmatrix}
				Q_{j} & S_{j}^{\prime}\\
				S_{j} & R_{j}
				\end{bmatrix}
				\begin{bmatrix}
				x_{j}\\
				u_{j}
			\end{bmatrix},
			\end{equation}
			\text{subject to $x_{0}=\boldsymbol{\chi}$,
                          $x_{N+1}=\boldsymbol{\zeta}$, and}
			\begin{align}
			%\begin{equation}
			0 &= {A}_j{x}_j + {B}_j{u}_j + {E}_j{x}_{j-1}
                            + {F}_j{x}_{j+1} +
                            {H}_j\boldsymbol{\xi}_j,~~j\in\mathcal{N}, \label{eq:stacked_system_dynamics}\\ 
			0 &= {C}_j{x}_j + {D}_j{u}_j + {\theta}_j -
                            \boldsymbol{\kappa}_j,~~j\in\mathcal{N},
                            \label{eq:stacked_inequality_constraints}\\   
			0 &\leq {\theta}_j,~~j\in\mathcal{N}, \label{eq:stacked_z_constraints}
			\end{align}
		\end{subequations}% for all $j\in\mathcal{N}$ 
		where
		%\small
		\begin{align*}
			{Q}_j &= \mathrm{blkdiag}(Q_{j,0}, \ldots,
                                Q_{j,T}) \in \mathbb{R}^{n_j(T\!+\!1)
                                \times n_j(T\!+\!1)},\\ 
			{R}_j &= \mathrm{blkdiag}(R_{j,0}, \ldots,
                                R_{j,T-1}) \in \mathbb{R}^{m_jT
                                \times m_jT},\\ 
			{S}_j &= [\mathrm{blkdiag}(S_{j,0}, \ldots,
                                S_{j,T-1})\ 0] \in \mathbb{R}^{m_jT
                                \times n_j(T\!+\!1)},\\ 
			{C}_j &= \mathrm{blkdiag}(C_{j,0}, \ldots,
                                C_{j,T}) \in
                                \mathbb{R}^{\nu_j(T\!+\!1) \times
                                n_j(T\!+\!1)},\\ 
			{D}_j &= [\mathrm{blkdiag}(D_{j,0}, \ldots,
                                D_{j,T-1})^{\prime}\ 0]^{\prime} \in
                                \mathbb{R}^{\nu_j(T\!+\!1) \times
                                m_jT},\\ 
			{H}_j &= [I\ 0\ \cdots\ 0]^{\prime} \in
                                \mathbb{R}^{n_j(T\!+\!1) \times
                                n_j},\\ 
			{\boldsymbol{\kappa}}_j &= \mathrm{col}(\boldsymbol{\kappa}_{j,0},
                                     \ldots, \boldsymbol{\kappa}_{j,T}) \in
                                     \mathbb{R}^{\nu_j(T\!+\!1)},\\ 
			{\boldsymbol{\chi}} &= \mathrm{col}({\boldsymbol{\chi}}_{0}, \ldots,
                                 {\boldsymbol{\chi}}_{T}) \in
                                 \mathbb{R}^{n_0(T\!+\!1)},\\ 
			{\boldsymbol{\zeta}} &= \mathrm{col}({\boldsymbol{\zeta}}_{0}, \ldots,
                                  {\boldsymbol{\zeta}}_{T}) \in
                                  \mathbb{R}^{n_{N\!+\!1}(T\!+\!1)},\\ 
			{A}_j &= 
			\renewcommand{\arraystretch}{1.2}
			\setlength{\arraycolsep}{1pt}
			\begin{bmatrix}
			\!-\!I      &       &          &\\
			A_{j,0} &\!-\!I     &          &\\
			&\ddots &\ddots    &\\
			& 		&A_{j,T\!-\!1} &\!-\!I
			\end{bmatrix},\ 
			{B}_j = 
			\renewcommand{\arraystretch}{1}
			\setlength{\arraycolsep}{1pt}
			\left[
			\begin{array}{ccc}
			0 		 &\cdots  	&0\\
			B_{j,0}  &\ddots	&\vdots \\
					 &\ddots 	&0\\
					 & 		 	&B_{j,T\!-\!1}
			\end{array}
			\right],\\
			{E}_j &= 
			\renewcommand{\arraystretch}{1}
			\setlength{\arraycolsep}{1pt}
			\begin{bmatrix}
			0       &       &          &\\
			E_{j,0} &0      &          &\\
			&\ddots &\ddots    &\\
			& 		&E_{j,T\!-\!1} &0
			\end{bmatrix}, \text{ and }
			{F}_j = 
			\renewcommand{\arraystretch}{1}
			\setlength{\arraycolsep}{1pt}
			\begin{bmatrix}
			0       &       &          &\\
			F_{j,0} &0      &          &\\
			&\ddots &\ddots    &\\
			& 		&F_{j,T\!-\!1} &0
			\end{bmatrix}.
		\end{align*}%
		Note that ${A}_j \in \mathbb{R}^{n_j(T\!+\!1) \times
                  n_j(T\!+\!1)}$,
                ${B}_j \in \mathbb{R}^{n_j(T\!+\!1) \times m_jT}$,
                ${E}_j \in \mathbb{R}^{n_j(T\!+\!1) \times
                  n_{j\!-\!1}(T\!+\!1)}$, and
                ${F}_j \in \mathbb{R}^{n_j(T\!+\!1) \times
                  n_{j\!+\!1}(T\!+\!1)}$. The block bi-diagonal
                structure of the matrices ${A}_j$ arises from the
                temporal structure of the system dynamics in
                the optimal control problem
                \eqref{eq:LQ_optimal_control_problem}.
		
		For the quadratic program
                \eqref{eq:temporal_stacked_problem}, the
                Karush-Kuhn-Tucker (KKT) conditions for optimality are
                given by
		\begin{subequations}\label{eq:kkt_conditions}
			\begin{align}
                          &{Q}_1{x}_1 + {S}_1^{\prime}{u}_1 + {A}_1^{\prime}{p}_1 + {C}_1^{\prime}{\lambda}_1 + {E}_{2}^{\prime}p_{2} = 0,\\
                          &{Q}_j{x}_j + {S}_j^{\prime}{u}_j + {A}_j^{\prime}{p}_j + {C}_j^{\prime}{\lambda}_j + {F}_{j\!-\!1}^{\prime}p_{j-1} \notag\\ 
                          &\qquad \qquad \qquad +~ {E}_{j\!+\!1}^{\prime}p_{j\!+\!1} = 0,\quad j \!\in\! \mathcal{N}\backslash\{1,N\},\\		
                          &{Q}_N{x}_N \!+\! {S}_N^{\prime}{u}_N \!+\! {A}_N^{\prime}{p}_N \!+\! {C}_N^{\prime}{\lambda}_N  + {F}_{N\!-\!1}^{\prime}p_{N\!-\!1} \!=\! 0,\\
                          &{S}_j{x}_j + {R}_j^{\prime}{u}_j + {B}_j^{\prime}{p}_j + {D}_j^{\prime}{\lambda}_j = 0,\quad j\in\mathcal{N},\\
                          &{A}_1{x}_1 + {B}_1u_1 + {E}_1\boldsymbol{\chi}  + {F}_1{x}_{2} + {H}_1\boldsymbol{\xi}_1\!=\! 0,\\
                          &{A}_j{x}_j + {B}_ju_j + {E}_j{x}_{j\!-\!1} +
                            {F}_j{x}_{j+1} + {H}_j\boldsymbol{\xi}_j = 0,\
                            \notag\\
                          &\qquad \qquad \qquad \qquad \qquad \qquad
                            \qquad \qquad
                            j\!\in\!\mathcal{N}\backslash\{1,N\},\\
                          &{A}_N{x}_N \!+\! {B}_Nu_N \!+\! {E}_N{x}_{N\!-\!1} \!+\! {F}_N\boldsymbol{\zeta} + {H}_N\boldsymbol{\xi}_N = 0,\\
                          &{C}_j{x}_j + {D}_j{u}_j \!-\! {\boldsymbol{\kappa}}_j + {\theta}_j = 0,\quad j\in\mathcal{N},\\
                          &{\Lambda}_j{\Theta}_j\mathbf{1} = 0,\quad \text{
                            and }\quad
                            %\ j\in\mathcal{N},\\
                            %&
                               [{\lambda}_j^{\prime}\
                               {\theta}_j^{\prime}]^{\prime} \geq 0,\quad
                               j\in\mathcal{N}, 
			\end{align}
		\end{subequations}
		where
                ${p}_j=\mathrm{col}(p_{j,0},\ldots,p_{j,T}) \in
                \mathbb{R}^{n_j(T\!+\!1)}$ and
                ${\lambda}_j=\mathrm{col}({\lambda}_{j,0}, \ldots,
                {\lambda}_{j,T}) \in \mathbb{R}^{\nu_j(T\!+\!1)}$ are
                Lagrange multipliers,
                ${\Lambda}_j = \mathrm{blkdiag}({\lambda}_{j,0},
                \ldots, {\lambda}_{j,T}) \in
                \mathbb{R}^{\nu_j(T\!+\!1)\times \nu_j(T\!+\!1)}$,
                ${\Theta}_j = \mathrm{blkdiag}({\theta}_{j,0}, \ldots,
                {\theta}_{j,T}) \in \mathbb{R}^{\nu_j(T\!+\!1)\times
                  \nu_j(T\!+\!1)}$, and $\mathbf{1}$ denotes a vector
                of all
                ones. %Given \eqref{eq:positive_semidefinite_cost},
                Since \eqref{eq:temporal_stacked_problem} is convex,
                the KKT conditions are necessary and sufficient for
                optimality~\cite{Nocedal2000numerical}.
			
		\subsection{Newton's Method}\label{subsec:newton_method}
                Various second-order optimization algorithms can be
                understood in terms of Newton's method for solving the
                KKT conditions (e.g., see
                \cite{Nocedal2000numerical}.) Typically, only a
                moderate number of Newton steps is required for
                convergence, and this is the main advantage over
                first-order optimization algorithms. The benefit comes
                from the use of second-order information, which can be
                constructed explicitly for quadratic programs. For
                the problem \eqref{eq:temporal_stacked_problem}, 
                the Newton steps in an interior point method (e.g.,
                see~\cite{Nocedal2000numerical}) take the form
		of the update	%
                \begin{equation}\label{eq:newton_step}
                  s^{(n + 1)} = s^{(n)} + \alpha^{(n)}\, \delta^{(n)},
                \end{equation}
                where $\alpha^{(n)}>0$ is a step size,
                $s^{(n)}~=~\mathrm{col}({s}_1^{(n)}, \ldots,
                {s}_N^{(n)})$,
                ${s}_j^{(n)}~=~\mathrm{col}({x}_j^{(n)},\
                {u}_j^{(n)},\ {p}_j^{(n)},\ {\lambda}_j^{(n)},\
                {\theta}_j^{(n)})$, and the second-order search
                direction
                $\delta^{(n)} = \mathrm{col}({\delta}_1^{(n)}, \ldots,
                {\delta}_N^{(n)})$ is obtained by solving the
                linearized KKT conditions, given by
                \begin{equation}\label{eq:kkt_equation}
                  \mathrm{blktrid}(\Phi^{(n)},\Omega)\,
                  \delta^{(n)} = b^{(n)}, 
                \end{equation}
                with $\Phi^{(n)}=(\Phi_j^{(n)})_{j\in\mathcal{N}}$,
                $\Omega=(\Omega_j)_{j\in\mathcal{N}\backslash\{1\}}$,
                $b^{(n)}=\mathrm{col}(b_1^{(n)},\ldots,b_N^{(n)})$,
                \begin{subequations}\label{eq:kkt_equations_submatrices_defined}
                  \begin{align}
                    \Phi_j^{(n)} &= 
                                   \begin{bmatrix}                         
                                     {Q}_j &{S}_j^{\prime}
                                     &{A}_j^{\prime}
                                     &{C}_j^{\prime} &0\\
                                     {S}_j &{R}_j &{B}_j^{\prime}
                                     &{D}_j^{\prime} &0\\
                                     {A}_j &{B}_j &0 &0 &0\\
                                     {C}_j &{D}_j &0 &0 &I\\
                                     0 &0 &0 &{\Theta}_j^{(n)}
                                     &{\Lambda}_j^{(n)}
                                   \end{bmatrix},\
                                       j \in
                                       \mathcal{N},
                                       \label{eq:Phi_j_kkt_equations_defined}\\ 
                    \Omega_j &=
                               \begin{bmatrix}
                                 0 &0 &{F}_{j-1}^{\prime} &0 &0\\
                                 0 &0 &0 &0 &0\\
                                 {E}_j &0 &0 &0 &0\\
                                 0 &0 &0 &0 &0\\
                                 0 &0 &0 &0 &0
                               \end{bmatrix},\ j \in
                               \mathcal{N}\backslash\{1\},
                               \label{eq:Omega_j_kkt_equations_defined} 
                    \\
                    % {\delta}_j^{(n)} &= \mathrm{col}
% 					\begin{pmatrix}
%        \delta_{{x}_j}^{(n)}, &\delta_{{u}_j}^{(n)},
%        &\delta_{{p}_j}^{(n)}, &\delta_{{\lambda}_j}^{(n)},
%        &\delta_{{\theta}_j}^{(n)}
% 					\end{pmatrix},\ j \in
% 					\mathcal{N},
%           \label{eq:delta_j_kkt_equations_defined}\\ 
%%
%%           b^{(n)}&=\mathrm{col}(b_1^{(n)},\ldots,b_N^{(n)}),
%%
% b_j^{(n)} &=~\mathrm{col}
% \begin{pmatrix}
% b_{{x}_j}^{(n)}, &b_{{u}_j}^{(n)}, &b_{{p}_j}^{(n)},
% &b_{{\lambda}_j}^{(n)}, &b_{{\theta}_j}^{(n)}
% \end{pmatrix},\ j \in \mathcal{N},
% \end{align}
% \text{with}
% \begin{align}
                            b_1^{(n)}&= \mathrm{col}(0,0,
                                       -{H}_1{\boldsymbol{\xi}}_1\!-\!
                                       E_1\boldsymbol{\chi}, 
                                       {\boldsymbol{\kappa}}_1,{\eta}_1^{(n)})
% \begin{pmatrix} 
% 0, &0, &-{H}_1{\boldsymbol{\xi}}_1\!-\! E_1\boldsymbol{\chi},
%  &{\boldsymbol{\kappa}}_1, &{\eta}_1^{(n)}	\end{pmatrix}   
%%\notag \\
%% &\phantom{=}\ 
\!-\!\Phi_1^{(n)}\!s_1^{(n)}\!\! -\!
     \Omega_{2}^{\prime}s_{2}^{(n)}\!\!, \label{eq:row_1_kkt_equation_defined}\\
  b_N^{(n)}&=
\mathrm{col}(0,0,-{H}_N{\boldsymbol{\xi}}_N\!-\!F_N\boldsymbol{\zeta},{\boldsymbol{\kappa}}_N,{\eta}_N^{(n)})   
% \mathrm{col}\begin{pmatrix}
% 0, &0, &-{H}_N{\boldsymbol{\xi}}_N\!-\! F_N\boldsymbol{\zeta},
% &{\boldsymbol{\kappa}}_N, &{\eta}_N^{(n)}
% \end{pmatrix}
                                   \notag \\
                                    &\phantom{=}\
                                       \!-\!\Omega_N
                                   s_{N\!-\!1}^{(n)}
                                   \!-\!\Phi_N^{(n)}s_N^{(n)}, \label{eq:row_N_kkt_equation_defined}\\
b_j^{(n)}&= 
\mathrm{col}(0,0,-{H}_j{\boldsymbol{\xi}}_j,{\boldsymbol{\kappa}}_j,{\eta}_j^{(n)})
                                        %            \mathrm{col}\begin{pmatrix}
					% 0, &0, &-{H}_j{\boldsymbol{\xi}}_j, &{\boldsymbol{\kappa}}_j, &{\eta}_j^{(n)}
					% \end{pmatrix}
  \notag \\
	&\phantom{=}\ -\Omega_j s_{j\!-\!1}^{(n)} \!-\! \Phi_j^{(n)}s_j^{(n)} \!\!-\!\Omega_{j\!+\!1}^{\prime}s_{j\!+\!1}^{(n)},\ j \in \mathcal{N}\backslash\!\{1,\!N\}, \label{eq:row_j_kkt_equation_defined}\\
% \intertext{and}
					{\eta}_j^{(n)} & = {\Theta}_j^{\!(n)\!}{\lambda_j}^{\!(n)\!} \!+\! {\Lambda}_j^{\!(n)\!}{\theta_j}^{\!(n)\!} 
					\!-\!
                                                         {\Lambda}_j^{\!(n)\!}{\Theta}_j^{\!(n)\!}\mathbf{1} \!+\! \sigma^{(n)\!}\mu^{(n)\!}\mathbf{1},\ j \in \mathcal{N}.\label{eq:etan}
				\end{align} 
                              \end{subequations}
                              In \eqref{eq:etan}, the scalar
                              $\mu^{(n)} \!=\!  \sum_{j=1}^{N}
                              (({\lambda}_j^{(n)})^{\prime}{\theta}_j^{(n)})/
                              \sum_{j=1}^{N}({\nu}_j(T\!+\!1))$ is a
                              measure of the duality gap and
                              $\sigma^{(n)} \in (0,1)$ is a centering
                              parameter. The
                              step-size scalar $\alpha^{(n)} > 0$ in
                              \eqref{eq:newton_step} is selected
                              (online) to ensure the components of
                              ${\lambda}_{j}^{(n + 1)}$ and
                              ${\theta}_{j}^{(n + 1)}$ remain
                              positive for 
                              $j\in\mathcal{N}$. The coefficient
                              matrix
                              $\mathrm{blktrid}(\Phi^{(n)},\Omega)$ in
                              \eqref{eq:kkt_equation} is non-singular
                              because, the matrices ${A}_j$ are
                              non-singular for all $j\in\mathcal{N}$
                              (see, \cite[Lemma
                              A.1]{Cantoni2017structured}.) 
			%%%%%%%%%%%%%%%%%% 
		\subsection{Structure-Preserving Block Elimination}\label{subsec:reduction_using_guassian_elimination}
                $\Lambda_j$, $\Theta_j$ and ${R}_j$ in
                \eqref{eq:Phi_j_kkt_equations_defined} are block
                diagonal, with block sizes that are independent of $N$ and
                $T$. For $j\in\mathcal{N}$, let
                ${\delta}_j^{(n)} = \mathrm{col}(\delta_{{x}_j}^{(n)},
                \delta_{{u}_j}^{(n)}, \delta_{{p}_j}^{(n)},
                \delta_{{\lambda}_j}^{(n)},
                \delta_{{\theta}_j}^{(n)})$ and
                $b_j^{(n)} = \mathrm{col}( b_{{x}_j}^{(n)},
                b_{{u}_j}^{(n)}, b_{{p}_j}^{(n)},
                b_{{\lambda}_j}^{(n)}, b_{{\theta}_j}^{(n)})$ be
                partitions aligned with the structure of
                $s_j^{(n)}$ noted below
                \eqref{eq:newton_step}. Dropping the Newton iteration index
                $(n)$, the ordered elimination of
	        \begin{align}
                  \delta_{\theta_j} &=
                                      {\Lambda}_j^{-1}(b_{\theta_j}
                                      -
                                      {\Theta_j}\delta_{\lambda_j}),  
                                      \label{eq:theta_elimination}\\    
                  \delta_{\lambda_j} &=
                                       -({\Theta}_j^{-1}{\Lambda}_j)
                                       (b_{\lambda_j} 
                                       -
                                       {C}_j\delta_{x_j}
                                       -{D}_j\delta_{u_j}
                                       -
                                       {\Lambda}_j^{-1}b_{\theta_j}),
                                       \label{eq:lambda_elimination}\\  
                  \delta_{u_j} &=
                                 \hat{R}_j^{-1}(\hat{b}_{u_j}
                                 - \hat{S}_j\delta_{x_j}
                                 -
                                 {B}_j^{\prime}\delta_{p_j}),
                                               \label{eq:u_elimination} 
                \end{align}
                from \eqref{eq:kkt_equation}, for $j \in \mathcal{N}$,
                yields the smaller symmetric
                % block tri-diagonal
                system
                % where $n_{\Delta} =
                % \sum_{j=1}^{N}(2{n}_j(T\!+\!1))$, can be
                % written as
                \begin{equation}\label{eq:reduced_kkt_equation}
                  \mathrm{blktrid}(\tilde{\Phi},\tilde{\Omega})\,
                  \tilde{\delta} = \tilde{b}, 
                \end{equation}
                where
                $\tilde{\Phi} = (\tilde{\Phi}_j)_{j\in\mathcal{N}}$,
                $\tilde{\Omega} =
                (\tilde{\Omega}_j)_{j\in\mathcal{N}\backslash\{1\}}$,
                $\tilde{\delta} =
                \mathrm{col}(\tilde{\delta}_1,\ldots,\tilde{\delta}_N)$,
                $\tilde{b} = \mathrm{col}(\tilde{b}_1,\ldots,\tilde{b}_N)$,
                and $\tilde{\delta}_j = \mathrm{col}(\delta_{{x}_j},
                \delta_{{p}_j}) \in \mathbb{R}^{2n_j(T+1)}$,
                $\tilde{b}_j = \mathrm{col}(\tilde{b}_{x_j},
                \tilde{b}_{p_j}) \in \mathbb{R}^{2n_j(T+1)}$,
                 \begin{subequations}
                   \label{eq:Phi_j_and_Omega_j_reduced_defined}
                   \begin{align}
                     \tilde{\Phi}_j &= 
                                      \begin{bmatrix}
					\tilde{Q}_j &\tilde{A}_j^{\prime}\\
					\tilde{A}_j &\tilde{R}_j
                                      \end{bmatrix} \in
                                      \mathbb{R}^{2n_j(T+1) \times
                                                      2n_j(T+1)},
                                    \label{eq:Phi_j_reduced_defined}\\ 
                     \tilde{\Omega}_j &= 
					\begin{bmatrix}
                                          0 &{F}_{j-1}^{\prime}\\
                                          {E}_j &0
					\end{bmatrix} \in
                                        \mathbb{R}^{2n_j(T+1) \times
                                          2n_j(T+1)},
                                      \label{eq:Omega_j_reduced_defined}\\ 
                     \begin{bmatrix}
                       \tilde{Q}_j &\tilde{A}_j^{\prime}\\
                       \tilde{A}_j &\tilde{R}_j
                     \end{bmatrix} &=
                                     \begin{bmatrix}
                                       \hat{Q}_j &{A}_j^{\prime}\\
                                       {A}_j &0
                                     \end{bmatrix} - \begin{bmatrix}
                                       \hat{S}_j^{\prime}\\
                                       {B}_j
                                     \end{bmatrix}\hat{R}_j^{-1}
                     \begin{bmatrix}
                       \hat{S}_j^{\prime}\\
                       {B}_j
                     \end{bmatrix}^{\prime}, \label{eq:cred}\\
                     \begin{bmatrix}
                       \hat{Q}_j &\hat{S}_j^{\prime}\\
                       \hat{S}_j &\hat{R}_j
                     \end{bmatrix} &=
                                     \begin{bmatrix}
                                       {Q}_j &{S}_j^{\prime}\\
                                       {S}_j &{R}_j
                                     \end{bmatrix} + \begin{bmatrix}
                                       {C}_j^{\prime}\\
                                       {D}_j^{\prime}
                                     \end{bmatrix}({\Theta}_j^{-1}{\Lambda}_j)
                     \begin{bmatrix}
                       {C}_j^{\prime}\\
                       {D}_j^{\prime}
                     \end{bmatrix}^{\prime}, \label{eq:dred}\\
                     % W_j &:= {\Lambda}_j^{-1}{\Theta}_j \succ 0,\\
                     \begin{bmatrix}
                       \tilde{b}_{x_j} \\
                       \tilde{b}_{p_j}
                     \end{bmatrix} &=
                                     \begin{bmatrix}
                                       \hat{b}_{x_j}\\
                                       \hat{b}_{p_j}
                                     \end{bmatrix} - \begin{bmatrix}
                                       \hat{S}_j^{\prime}\\
                                       {B}_j
                                     \end{bmatrix}\hat{R}_j^{-1}\hat{b}_{u_j},
                     \label{eq:ered}\\
                     \begin{bmatrix}
                       \hat{b}_{x_j} \\
                       \hat{b}_{u_j}
                     \end{bmatrix} &=
                                     \begin{bmatrix}
                                       {b}_{x_j}\\
                                       {b}_{u_j}
                                     \end{bmatrix} +
                     \begin{bmatrix}
                       {C}_j^{\prime} & {C}_j^{\prime}\\
                       {D}_j^{\prime} & {D}_j^{\prime}
                     \end{bmatrix}
                                        \begin{bmatrix}
                                          ({\Theta}_j^{-1}{\Lambda}_j)
                                          b_{\lambda_j}\\
                                          {\Theta}_j^{-1}b_{\theta_j}
					\end{bmatrix}, \label{eq:fred}
                   \end{align}
                 \end{subequations}
                 for $j \in \mathcal{N}$. Note that the computations
                 required to form \eqref{eq:cred}--\eqref{eq:fred} are
                 decomposable. Only the manipulation of block diagonal
                 matrices, with block sizes independent of $N$ and
                 $T$, is required. Moreover, the structure of
                 \eqref{eq:kkt_equation} is preserved in
                 \eqref{eq:reduced_kkt_equation}.  Further, it is of
                 note that in \eqref{eq:Phi_j_reduced_defined},
                 $\tilde{Q}_j$ and $\tilde{R}_j$ are block diagonal,
                 and $\tilde{A}_j$ is block bi-diagonal.

                 Next, an iterative algorithm based on the PCG method
                 is developed to solve
                 \eqref{eq:reduced_kkt_equation}. The number of
                 iterations required depends on the quality of the
                 preconditioner used. In the worst case, the maximum
                 number of iterations is $O(NT)$, i.e., the size of
                 the problem. The worst case computational complexity
                 of the proposed approach is thus $O(N^2T^2)$, since
                 the computational complexity of each PCG step is
                 shown to be $O(NT)$ for the structured problem at
                 hand. This is (at the least) no worse than the previously
                 discussed direct methods when $T\approx N$. However,
                 good preconditioning can substantially reduce the
                 number of PCG iterations needed. The properties of a
                 structured preconditioner are detailed in
                 Section~\ref{sec:structured_preconditioner_for_CG_method}.
                 This is the main contributions of the work.
			%%%% End of section
	\section{PCG Solvers}\label{sec:pcg_method}
        The conjugate gradient (CG) method is an iterative Krylov subspace
        method. It is used for solving linear systems
        of equations with positive-definite coefficient
        matrix~\cite{Hestenes1952methods}. While non-singular, the
        block tri-diagonal matrix
        $\mathrm{blktrid}(\tilde{\Phi},\tilde{\Omega})$ in
        \eqref{eq:reduced_kkt_equation} has both positive and negative
        eigenvalues. This indefinite system can be solved using other
        Krylov methods, like MINRES \cite{Paige1975solution} or GMRES
        \cite{Saad1986gmres}. However, the computations for these are
        more involved than the CG method, with reduced scope for
        decomposability in the case of structured
        problems. Transforming both sides of
        \eqref{eq:reduced_kkt_equation} by
        $\mathrm{blktrid}(\tilde{\Phi},\tilde{\Omega})$ from the left
        yields the positive-definite system of equations
        \begin{equation}
          \label{eq:block_pentadiagonal_PD_reduced_kkt_equation} 
          \Psi \tilde{\delta} = \breve{b},
        \end{equation}
        where
        $\Psi = (\mathrm{blktrid}(\tilde{\Phi},\tilde{\Omega}))^2$
        and
        $\breve{b} = \mathrm{blktrid}(\tilde{\Phi},\tilde{\Omega})\,
        \tilde{b}$.
        %Since,
        %$\mathrm{blktrid}(\tilde{\Phi},\tilde{\Omega})$ is block
        %tri-diagonal, 
        The positive-definite matrix $\Psi$ is now block
        penta-diagonal, but 
        \eqref{eq:block_pentadiagonal_PD_reduced_kkt_equation} now is
        amenable to the CG method.

%        Now, the CG method can be used to solve
%        \eqref{eq:block_pentadiagonal_PD_reduced_kkt_equation}. 
        Let $e^{(i)} = \tilde{\delta}^{(i)} - \tilde{\delta}^{*}$
        be the error between $i$-th iterate $\tilde{\delta}^{(i)}$
        of the CG method and the exact solution
        $\tilde{\delta}^*$ of
        \eqref{eq:block_pentadiagonal_PD_reduced_kkt_equation}. It can
        be shown that $e^{(i)}$ satisfies the
        following~\cite[Thm.~6.29]{Saad2003iterative}:  
		\begin{equation}\label{eq:convergence_rate_CG_method}
			\|e^{(i)}\|_{\Psi} \leq
                        2\left((\sqrt{\kappa(\Psi)}
                            -1)\big/ (\sqrt{\kappa(\Psi)} +
                            1)\right)^{i}\|e^{(0)}\|_{\Psi}, 
		\end{equation}
		where $\|e\|_{\Psi} = e^{\prime}\Psi e$,
                $\kappa(\Psi) = \lambda_{\mathrm{max}}(\Psi)\big/
                \lambda_{\mathrm{min}}(\Psi)$ is the condition number,
                and $\lambda_{\mathrm{max}}(\Psi)$
                (resp.~$\lambda_{\mathrm{min}}(\Psi)$) is the maximum
                (resp.~minimum) eigenvalue of $\Psi$. As such, the CG
                method converges faster for $\kappa(\Psi)$ closer to
                $1$.
                %However, for ill-conditioned matrices (i.e.,
                %$\kappa(\Psi)\gg 1$), the convergence can be very
                %slow.
                To improve the condition number,
                problem
                \eqref{eq:block_pentadiagonal_PD_reduced_kkt_equation}
                can be transformed into 
		\begin{equation}\label{eq:pcg_system}
			P^{-1/2}\Psi P^{-1/2} \breve{\delta} =
                        P^{-1/2}\breve{b}, 
		\end{equation}
		where $\breve{\delta} = P^{1/2}\tilde{\delta}$ and
                $P=P^\prime\succ 0$. The CG method is then applied to
                \eqref{eq:pcg_system}.
                %This is known as the proposed
                %PCG method.
                % \cite{Axelsson1974preconditioning}.
                An efficient implementation of this PCG (i.e., preconditioned
                CG) method is given in
                Algorithm~\ref{alg:pcg}~\cite{Hackbusch2016iterative}.
                %from~\cite{Concus1976generalized},
                %~\cite{Hackbusch2016iterative}.  
		%
		\begin{algorithm}[t]
			\caption{PCG
                          for
                          \eqref{eq:block_pentadiagonal_PD_reduced_kkt_equation}
                          with preconditioner
                          $P$.}  
			\label{alg:pcg}
			\begin{algorithmic}[1] % [1] is for line numbers
				\State \textbf{Initialize}
                                $\tilde{\delta}^{(0)}$, $\epsilon$,
                                $\text{iter}_{\text{max}}$ 
				\State $r^{(0)} = \breve{b} - \Psi
                                \tilde{\delta}^{(0)}$ 
				\State \textbf{Solve} $P d^{(0)} =
                                r^{(0)}$ \label{step:precond0}
				\State $\beta^{(0)} = 
                                (d^{(0)})^\prime r^{(0)} $  
				\State Set $i = 0$ \label{step:initialize}
				\While {$i < \text{iter}_{\text{max}}$}
				\State $y^{(i)} = \Psi
                                {d}^{(i)}$ \label{step:matrix_vector_product} 
				\State $\gamma^{(i)} =
                                {\beta^{(i)}}/{(
                                  (y^{(i)})^\prime d^{(i)})
                                }$ \label{step:alpha_update} 
				\State $\tilde{\delta}^{(i+1)} =
                                \tilde{\delta}^{(i)} 
                                + \gamma^{(i)}
                                d^{(i)}$\label{step:variable_update} 
				\State $r^{(i+1)} = r^{(i)} -
                                \gamma^{(i)}
                                y^{(i)}$\label{step:residual_update} 
				\State \textbf{if $\|r^{(i+1)}
                                  \|_{\infty} < \epsilon $
                                  exit}  \label{step:error_update} 
				\State \textbf{Solve} $P q^{(i+1)} =
                                r^{(i+1)}$ \label{step:preconditioning} 
				\State $\beta^{(i+1)} = 
                                (q^{(i+1)})^\prime r^{(i+1)}
                                $  \label{step:beta_update} 
				\State $d^{(i+1)} = r^{(i+1)} +
                                \left({\beta^{(i+1)}}/{\beta^{(i)}}\right)
                                d^{(i)}$ \label{step:direction_update} 
				\State $i = i+1$
				\EndWhile
                                %\\
			%	Result ${\delta}^{(i+1)}$ \label{step:result}
			\end{algorithmic}
		\end{algorithm}

                The preconditioner $P=\Psi$ would give
                $P^{-1/2}\Psi P^{-1/2} = I$. But
                steps~\ref{step:precond0}
                and~\ref{step:preconditioning} of
                Algorithm~\ref{alg:pcg} are then the original
                problem. Incomplete sparse LU factorization of $\Psi$
                can be used for $P$ instead. Such preconditioners are
                considered in \cite{Benzi2003robust,
                  Xia2017effective}.  However, for the resulting
                preconditioner to be positive definite and effective,
                it may be necessary to use incomplete LU factors that
                are denser (i.e., have less structure) than $\Psi$.
		
		In the next two sections, a structured
                approach is developed for the preconditioning
                steps. Specifically, it is proposed to use a fixed
                number of block Jacobi iterations (e.g., see
                \cite{Hackbusch2016iterative}) to {\em
                  approximately} solve steps~\ref{step:precond0}
                and~\ref{step:preconditioning} with $P=\Psi$. The
                approach builds on ideas borrowed from
                \cite{Concus1976generalized, Johnson1983polynomial,
                  Adams1985mstep}.    
	%%%%%% end of section
	\section{Block Jacobi
          Preconditioning}\label{sec:structured_preconditioner_for_CG_method} 
        Let $\mathcal{K} = \{1,2,\ldots, K\}$, with $K=\lceil N/2 \rceil$, i.e., 
        $K=N/2$ when $N$ is even, and $K=(N+1)/2$ otherwise.
        Also define
		\begin{subequations}
                  \label{eq:constructing_Pi_ell_Upsilon_g_matrices} 
			\begin{equation}\label{eq:structure_Pi_ell_2x2_eq_a}
			\Delta_k = 
			\begin{bmatrix}
				Z_{2k - 1} &Y_{2k}^{\prime}\\
				Y_{2k}	 &Z_{2k}
			\end{bmatrix},~ k \in \mathcal{K}\backslash\{K\}, 
			\end{equation}
			\begin{equation}\label{eq:structure_Pi_ell_2x2_eq_b}
			\Delta_{K} =	
			\begin{cases}
				\begin{bmatrix}
					Z_{N-1} &Y_{N}^{\prime}\\
					Y_{N}	 &Z_{N}
				\end{bmatrix}, & \text{$N$ even},\\
				Z_N, & \text{$N$ odd},
			\end{cases}
			\end{equation}
%			\text{and}
			\begin{equation}
				\Upsilon_k = 
				\begin{bmatrix}
					V_{2k - 1} &Y_{2k - 1}\\
					0		 &V_{2k}
				\end{bmatrix},~ k \in
                                \mathcal{K}\backslash\{1,K\},  
			\end{equation}
			\begin{equation}
			\Upsilon_{K} =	
			\begin{cases}
				\begin{bmatrix}
					V_{N-1} &Y_{N-1}\\
					0		 &V_{N}
				\end{bmatrix}, & \text{$N$ even},\\ 
				\begin{bmatrix}
					V_{N} &Y_{N}
				\end{bmatrix}, & \text{$N$ odd},
			\end{cases}
			\end{equation}
                      \end{subequations}
                       %with $Y_1 = 0$, $V_1 = 0$ and $V_2 = 0$,
		where referring to
                \eqref{eq:Phi_j_and_Omega_j_reduced_defined}, 
		\begin{subequations}\label{eq:blocks_of_Delta_square_defined}
			\begin{align}
				Z_j &= \tilde{\Phi}_j^2 +
                                      \tilde{\Omega}_j
                                      \tilde{\Omega}_j^{\prime} +
                                      \tilde{\Omega}_{j+1}^{\prime}
                                      \tilde{\Omega}_{j+1},
                                      \label{eq:Z_j_defined}\\    
				Y_j &=
                                      \tilde{\Omega}_{j}\tilde{\Phi}_{j-1}
                                      +
                                      \tilde{\Phi}_{j}\tilde{\Omega}_{j},
                                      \label{eq:Y_j_defined}\\     
				V_j &=
                                      \tilde{\Omega}_{j}\tilde{\Omega}_{j-1},
                                      \label{eq:V_j_defined}
			\end{align}
		\end{subequations}
		for $j \in \mathcal{N}$, with 
                $\tilde{\Omega}_{N+1}=0$.
		Given this,
                $\Psi = \mathrm{blktrid}(\varDelta,\Upsilon)$, where
                $\varDelta=(\Delta_k)_{k\in\mathcal{K}}$ and
                $\Upsilon=(\Upsilon_k)_{k\in\mathcal{K}\backslash\{1\}}$.
                Moreover, the preconditioning steps~\ref{step:precond0}
                and~\ref{step:preconditioning} with $P=\Psi$,
                      %\eqref{eq:block_pentadiagonal_PD_reduced_kkt_equation}
                      can be re-written in the form
		\begin{equation}
                  \label{eq:block_tridiagonal_PD_reduced_kkt_equation} 
			\mathrm{blktrid}(\varDelta,\Upsilon)\, \zeta = \tau.
		\end{equation}
		 
				Let
                $\Delta=\mathrm{blkdiag}(\Delta_1,\ \ldots,\
                \Delta_{K})$ and
                $\Sigma = \Delta - \Psi$.
		
                The block Jacobi method for solving 
                \eqref{eq:block_tridiagonal_PD_reduced_kkt_equation}
                involves the following iterations:
		\begin{equation}\label{eq:jacobi_iterations}
			\Delta\zeta^{(l+1)} = \tau + \Sigma\zeta^{(l)}.
		\end{equation}
	        Since $\Psi = \mathrm{blktrid}(\varDelta,\Upsilon)
                \succ 0$ is block tri-diagonal, it is known that these
                iterations converge \cite{Hackbusch2016iterative}. The
                proposal is to apply just a fixed number of
                Jacobi iterations for the preconditioning steps of
                Algorithm~\ref{alg:pcg}.
                Characteristics of this approach are
                discussed in the next three sub-sections.
		\subsection{Positive definiteness of the
                  preconditioner}  
		Executing a fixed number of block Jacobi steps from
                zero is equivalent to the use of a positive-definite
                preconditioner.
                %This result is established in the following
                %theorem.
		\begin{thm}\label{thm:Positive_definiteness_of_preconditioner}
                  Given $L\in\mathbb{N}$ and $\zeta^{(0)}=0$, the
                  $L$-th iterate of
                  \eqref{eq:jacobi_iterations} satisfies
                  $P_L \zeta^{(L)} = \tau$ with $P_L = W_L^{-1}$, where
                  $W_L=\sum_{l=0}^{L-1}(\Delta^{-1}\Sigma)^l\Delta^{-1}
                  \succ 0$.
		\end{thm}
		\begin{IEEEproof}
                  Noting that $\Delta \succ 0$ is invertible, it
                  follows from \eqref{eq:jacobi_iterations} that
                  $\zeta^{(L)}~=~W_L\tau + (\Delta^{-1}\Sigma)^L
                  \zeta^{(0)} = W_L
                  \tau$. It is established below that $W_L$ is 
                  positive definite, and thus, invertible. As such,
                  $P_L \zeta^{(L)} = W_L^{-1} \zeta^{(L)} = \tau$.
			
                  Positive definiteness of $W_L$ is a consequence of
                  the known property
                  $\Psi = \mathrm{blktrid}(\Delta,\Upsilon) \succ
                  0$. With $U = \mathrm{blkdiag}(U_{K},...,U_1)$, and
                  $U_k = (-I)^k$ for $k=1,\ldots,K$, first note that
                  $2\Delta - \Psi = U^{\prime} \Psi U \succ 0$. Then
                  note that $W_1 = \Delta^{-1} \succ 0$, and using
                  $(\Delta^{-1}\Sigma)^l\Delta^{-1}=
                  \Delta^{-1}(\Sigma\Delta^{-1})^l$, 
                  that 
                  \begin{align*}
                    W_{2M} &= \sum_{l=0}^{M-1}
                             (\Delta^{-1}\Sigma)^l \Delta^{-1}
                             (\Delta + \Sigma) \Delta^{-1} (\Sigma
                             \Delta^{-1})^l  \\ 
                           &= \sum_{l=1}^{M-1}  (\Delta^{-1}\Sigma)^l
                             \Delta^{-1} (2\Delta - \Psi) \Delta^{-1} (\Sigma
                             \Delta^{-1})^l \\ 
                           &\quad + \Delta^{-1} (2\Delta - \Psi)
                             \Delta^{-1}\succ 0, 
                  \end{align*}
                  and
                  \begin{align*}
                    W_{2M+1} &= \sum_{l=0}^{2M-1}
                               (\Delta^{-1}\Sigma)^l\Delta^{-1} +
                               (\Delta^{-1}\Sigma)^{2M}\Delta^{-1}  
                    \\ 
                             &= \sum_{l=0}^{M-1} (\Delta^{-1}\Sigma)^l
                               \Delta^{-1} (\Delta + \Sigma) \Delta^{-1} (\Sigma
                               \Delta^{-1})^i\\  
                             &\quad + (\Delta^{-1}\Sigma)^M \Delta^{-1}
                               ((\Delta^{-1}\Sigma)^M)^{\prime}\succ 0,
                  \end{align*}
                  for
                  $M\in\mathbb{N}$. Therefore, $W_L\succ 0$, as claimed.
		\end{IEEEproof}
		\subsection{An analytic bound on achieved
                  conditioning}
                \label{subsec:effectiveness_of_block_Jacobi_preconditioner}
                The iterations \eqref{eq:jacobi_iterations} converge
                to the solution of
                \eqref{eq:block_tridiagonal_PD_reduced_kkt_equation}
                if and
                only if
                \begin{equation}\label{eq:convergence_condition}
                  \varrho(\Delta^{-1}\Sigma) < 1,
                \end{equation}
                where $\varrho(\cdot)$ denotes spectral
                radius~\cite[Thm 2.16]{Hackbusch2016iterative}. For
                $\Psi = \mathrm{blktrid}(\varDelta,\Upsilon) \succ 0$,
                and the split $\Psi=\Delta-\Sigma$, condition
                \eqref{eq:convergence_condition} holds~\cite[Lem 4.7, 
                Thm. 4.18]{Hackbusch2016iterative}.
                \begin{thm}\label{thm:spectrum_of_preconditioned_system}
                  With
                  $P_L=(\sum_{l=0}^{L-1}(\Delta^{-1}\Sigma)^l\Delta^{-1})^{-1}$
                  for given $L\in\mathbb{N}$,
                  % The spectral condition number $\kappa$ of the
                  % preconditioned system $(P_L^{-1/2}\Psi P_L^{-1/2})$
                  % is given as
                  \begin{equation}
                    \label{eq:condition_number_of_preconditioned_system}
                    \kappa(P_L^{-1/2}\Psi P_L^{-1/2}) \leq \frac{1 +
                      (\varrho(\Delta^{-1}\Sigma))^L}{1 -
                      (\varrho(\Delta^{-1}\Sigma))^L}.  
                  \end{equation}	
                \end{thm}
                \begin{IEEEproof}
                  By
                  Theorem~\ref{thm:Positive_definiteness_of_preconditioner},
                  $P_L\succ 0$. Using $\Psi=\Delta-\Sigma$,
                  \begin{align}
                    P_L^{-1}\Psi
                    % &= P_m(\Delta - \Sigma)
                    % &=
                    %   \sum_{l=0}^{L-1}(\Delta^{-1}\Sigma)^{l}\Delta^{-1}(\Delta
                    %   - \Sigma)\notag\\ 
                      &= \sum_{l=0}^{L-1}(\Delta^{-1}\Sigma)^{l}(I -
                        \Delta^{-1}\Sigma)
                        %\notag\\ 
                      % &= (I + \Delta^{-1}\Sigma + \cdots +
                      %   (\Delta^{-1}\Sigma)^{l-1})(I -
                      %   \Delta^{-1}\Sigma)\notag\\ 
                      %   &
                            = I - (\Delta^{-1}\Sigma)^{L}.
                        \label{eq:spectrum_preconditioned_system}
                  \end{align}
                  Furthermore,
                  $P_L^{-1}\Psi=P_L^{-1/2}(P_L^{-1/2}\Psi
                  P_L^{-1/2})P_L^{1/2}$, whereby
                  $\mathrm{spec}(P_L^{-1/2}\Psi P_L^{-1/2}) =
                  \mathrm{spec}(P_L^{-1}\Psi)$.
%                  , where
%                  $\mathrm{spec}(\cdot)$ denotes the set of
                            %                             eigenvalues.
                  So the result holds as
                  $\lambda_{\text{max}}(I-(\Delta^{-1}\Sigma)^L) \leq
                  1+(\varrho(\Delta^{-1}\Sigma))^L$ and
                  $\lambda_{\text{min}}(I-(\Delta^{-1}\Sigma)^L)\geq
                  1-(\varrho(\Delta^{-1}\Sigma))^L>0$.
                \end{IEEEproof}

                By
                Theorem~\ref{thm:spectrum_of_preconditioned_system},
                the number $L$ of block Jacobi iterations can be
                selected to achieve desired conditioning.
		%%% end of subsection
		\subsection{Decomposable computations}
                \label{subsec:computation_of_Jacobi_preconditioner}
                Note that explicit construction of the preconditioner
                $P_L$ is not needed. At each PCG iteration, $L$
                iterations of \eqref{eq:jacobi_iterations} are
                performed from $\zeta^{(0)}=0$.
                Since $\Delta$ is block diagonal, the computations
                required to implement each Jacobi iteration can be
                decomposed into $K=\lceil N/2 \rceil$ smaller problems
                \begin{equation}
                  \label{eq:small_block_Jacobi_linear_system_with_Pi_ell}
                  \Delta_{k} \zeta_{k}^{(l+1)} = \omega_{k}, 
                \end{equation} 
                where
                $\omega_{k} = \tau_{k} + \Upsilon_{k}\zeta_{k-1}^{(l)}
                + \Upsilon_{k+1}^{\prime}\zeta_{k+1}^{(l)}$ for
                $k \in \mathcal{K}$, with $\Upsilon_{K+1}=0$.
                %Hence,
                %the computational complexity scales linearly with
                %$K$.
                Each $\Delta_{k}$ is a block $2\times2$ matrix,
                with inner blocks that are
                structured. To see this structure, consider
                \begin{equation}
                  \Delta_k =
                  \begin{bmatrix}
                    Z_{2k-1} &Y_{2k}^{\prime}\\
                    Y_{2k} &Z_{2k}
                  \end{bmatrix}.
                \end{equation}
                % The structure of all $\Delta_k$ is essentially the
                % same.
                Note that
                \begin{subequations}\label{eq:subblock_Pi}
                  \begin{align}
                    Z_{j} &= \tilde{\Phi}_{j}^2 
                            + \tilde{\Omega}_j\tilde{\Omega}_j^\prime
                            + \tilde{\Omega}_{j+1}^{\prime}\tilde{\Omega}_{j+1}
                            \notag\\ 
                          &=\left[\!\begin{smallmatrix}
                            \tilde{Q}_j^2\!+\!
                            \tilde{A}_j^{\prime}\tilde{A}_j\!+\!
                            F_{j\!-\!1}F_{j\!-\!1}^\prime \!+\! E_j^{\prime}E_j
                            &\tilde{Q}_j\tilde{A}_j^{\prime} \!+\!
                            \tilde{A}_j^{\prime}\tilde{R}_j\\ 
                            \tilde{A}_j\tilde{Q}_j+\tilde{R}_j\tilde{A}_j
                            &\tilde{A}_j\tilde{A}_j^{\prime}\!+\!
                            \tilde{R}_j^2 \!+\!
                            F_jF_j^{\prime}\!+\!E_{j\!+\!1}^\prime E_{j\!+\!1}
                          \end{smallmatrix}\!\right],\\
                    Y_j &= \tilde{\Omega}_{j}\tilde{\Phi}_{j\!-\!1} \!+\!
                          \tilde{\Phi}_{j}\tilde{\Omega}_{j}
                          %,\notag\\ 
                          %&
                            \!=\!\left[\!\begin{smallmatrix}
                            F_{j\!-\!1}^{\prime}\tilde{A}_{j\!-\!1} \!+\!
                            \tilde{A}_j^{\prime}E_j &
                            F_{j\!-\!1}^{\prime}\tilde{R}_{j\!-\!1} \!+\! \tilde{Q}_j
                            F_{j\!-\!1}^{\prime}\\ 
                            E_j\tilde{Q}_{j\!-\!1} \!+\! \tilde{R}_{j\!-\!1}E_j
                            &E_j\tilde{A}_{j\!-\!1}^{\prime} \!+\!
                            \tilde{A}_jF_{j\!-\!1}^{\prime} 
                          \end{smallmatrix}\!\right]\!.
                  \end{align}
                \end{subequations}
			%The structure of $Z_2$ is similar to that of
                        %$Z_1$.
                All blocks components of \eqref{eq:subblock_Pi} are
                block diagonal, except for the block bi-diagonal
                $\tilde{A}_j$ for $j\in\mathcal{N}$. The sub-block
                sizes are all independent of both $N$ and $T$. The
                diagonal blocks of $Z_j$ are block tri-diagonal, while
                off-diagonal blocks are block bi-diagonal for
                $j \in \mathcal{N}$. Similarly, the diagonal blocks of
                $Y_j$ are block tri-diagonal, and the off-diagonal
                blocks are block diagonal for
                $j \in \mathcal{N}\backslash\{1\}$. To summarize, the
                matrices $\Delta_{k}$ have block-banded structure.
                In particular, there exists a permutation of variables
                such that 
                \eqref{eq:small_block_Jacobi_linear_system_with_Pi_ell}
                takes the form
                \begin{equation}
                  \label{eq:small_block_tridiagonal_Jacobi_linear_system_with_Pi_ell}
                  \mathrm{blktrid}(\Xi_{k},\varPi_{k})\,
                  \hat{\zeta}_{k}^{(l+1)} = \hat{\omega}_{k}, 
                \end{equation}
                where
                $\hat{\omega}_{k} = \hat{\tau}_{k} +
                \hat{\Upsilon}_{k}\hat{\zeta}_{k-1}^{(i)} +
                \hat{\Upsilon}_{k+1}^{\prime}\hat{\zeta}_{k+1}^{(i)}$,
                $\Xi_k = (\Xi_{k,t})_{t\in\mathcal{T}}$, $\varPi_k =
                (\varPi_{k,t})_{t\in\mathcal{T}\backslash\{T\}}$,
                \begin{subequations}
                  \label{eq:structure_of_diagonal_blocks_of_tridiagonal_Pi_ell}
                  \begin{equation}\label{eq:structure_of_small_pi_gt}
                    \Xi_{k,t} =
                    \setlength{\arraycolsep}{3pt} 
                    \left[
                      \begin{array}{cccc}
                        \check{Q}_{2k-1,t} & \varOmega_{2k-1,t}
                        &\check{E}_{2k,t}^{\prime} & 0\\  
                        \varOmega_{2k-1,t} & \check{R}_{2k-1,t-1} & 0
                                                   & \check{F}_{2k,t-1}^{\prime}\\ 
                        \check{E}_{2k,t} &0 &\check{Q}_{2k,t} &
                                                                \varOmega_{2k,t}\\  
                        0 &\check{F}_{2k,t-1} &\varOmega_{2k,t}
                                                   &\check{R}_{2k,t-1}   
                      \end{array}
                    \right],
                  \end{equation}
                  %for $k \in \mathcal{K}$ and
                  %  $t \in \mathcal{T}$, 
                  with
                  \begin{align}
                    \check{Q}_{j,t} &\!=\! \tilde{Q}_{j,t}^2 \!+\! I \!+\! \tilde{A}_{j,t}^{\prime}\tilde{A}_{j,t} \!+\! E_{j,t}^{\prime}E_{j,t} \!+\! F_{j-1,t}^{\prime}F_{j-1,t},\\
                    \check{R}_{j,t} &\!=\! \tilde{R}_{j,t}^2 \!+\! I \!+\! \tilde{A}_{j,t}\tilde{A}_{j,t}^{\prime} \!+\! E_{j,t}E_{j,t}^{\prime} \!+\! F_{j,t}F_{j,t}^{\prime},\\
                    \varOmega_{j,t} &\!=\! -\tilde{Q}_{j,t} \!-\! \tilde{R}_{j,t-1},\\
                    \check{E}_{j,t} &\!=\! \tilde{A}_{j,t}^{\prime}E_{j,t} \!+\! F_{j-1,t}^{\prime}\tilde{A}_{j-1,t},\\
                    \check{F}_{j,t} &\!=\! \tilde{A}_{j,t}F_{j-1,t}^{\prime} \!+\! E_{j,t}\tilde{A}_{j-1,t}^{\prime},% \\
                    % \text{for } & j \in \mathcal{N}, \text{ and } t \in \mathcal{T},\notag
                  \end{align}
                \end{subequations}
                for $j\in\mathcal{N}$, and
                \begin{subequations}
                  \begin{equation}\label{eq:structure_of_small_phi_gt}
				\varPi_{k,t} =
				\setlength{\arraycolsep}{1pt} 
				\left[
				\begin{array}{cccc}
				-\tilde{A}_{2k-1,t} &0
                                  &-F_{2k-1,t}^{\prime} &0\\
				\check{A}_{2k-1,t}
                                                    &-\tilde{A}_{2k-1,t}
                                  & G_{2k-1,t} &-F_{2k-1,t}^{\prime}\\
				-E_{2k,t} &0 & -\tilde{A}_{2k,t} &0\\
				X_{2k,t} &-E_{2k,t}
                                  &\check{A}_{2k,t} &
                                                      -\tilde{A}_{2k,t} 
				\end{array}
                              \right]
                            \end{equation}
                            with
			%	\text{for all $k \in
                        %	\mathcal{K}\backslash\{1\}$ and $t \in
                        %	\mathcal{T}\backslash\{T\}$ with}
                            \begin{align}
                              \check{A}_{j,t} &=
                                                \tilde{A}_{j,t}\tilde{Q}_{j,t}
                                                +
                                                \tilde{R}_{j,t}\tilde{A}_{j,t},~~
                                                j\in\mathcal{N}\\
                              G_{j,t} &=
                                        F_{j-1,t}^{\prime}\tilde{R}_{j-1,t} +
                                        \tilde{Q}_{j,t}F_{j-1,t}^{\prime},~~
                              j\in\mathcal{N}\backslash\{1\}\\
                              X_{j,t} &= E_{j,t}\tilde{Q}_{j-1,t} +
                                        \tilde{R}_{j-1,t}E_{j,t},~~
                                        j\in\mathcal{N}\backslash\{1\}. 
                            \end{align}
                          \end{subequations}
                          Note that
                          $\Xi_{k,t} , \varPi_{k,t} \in
                          \mathbb{R}^{\hat{n}_{k,t}\times
                          \hat{n}_{k,t}}$, where
                        $\hat{n}_{{k,t}}=2(n_{2k - 1}+n_{2k})$ for all
                        $k \in \mathcal{K}$ and $t \in
                        \mathcal{T}$. That is, the sizes of the sub-blocks
                        of $\mathrm{blktrid}(\Xi_{k},\varPi_{k})$ are
                        independent of $N$ and $T$.

                        For each $k\in\mathcal{K}$, the block
                        tri-diagonal system
                        \eqref{eq:small_block_tridiagonal_Jacobi_linear_system_with_Pi_ell}
                        can be solved by backward-forward recursions,
                        with computational complexity $O(2T)$, that
                        effectively implement an LDL factorization
                        method~\cite{Meurant1992review}. In this way,
                        the preconditioning computations decompose
                        into a collection of $\lceil N/2 \rceil$
                        parallel threads each comprising computations
                        for $2T$ sequential (possibly dense) problems
                        of size that is independent of $N$ and $T$.
                        Table~\ref{tab:PCG_complexity_analysis}
                        provides a complexity analysis of each step of
                        Algorithm~\ref{alg:pcg}, including the
                        inter-thread data exchange overhead for an
                        implementation with parallelism.
                        %for both
                        %implementations.
                        
                        \renewcommand{\arraystretch}{1.5}
			\begin{table}[t]
				\begin{tabular}{l|l|ll}
                                  \hline
                                  \rowcolor[HTML]{FFCE93} 
                                  &
                                    \textbf{\begin{tabular}[c]{@{}l@{}}Single
                                              Thread \end{tabular}} &
                                                                      \multicolumn{2}{l}{\cellcolor[HTML]{FFCE93}\textbf{\begin{tabular}[c]{@{}l@{}}N/2 Parallel Threads\end{tabular}}} \\ \hline
                                  \rowcolor[HTML]{ECF4FF} 
                                  \textbf{PCG Steps}
                                  & \textbf{Computations}
                                   & \textbf{\begin{tabular}{l}
                                               \!\!\!\!Computations\\[-2pt]
                                               \!\!\!\!\!\! per
                                               thread\end{tabular}}
                                  &
                                    \textbf{\begin{tabular}{l}\!\!\!\!
                                              Data
                                              xchg.\\[-2pt]  
                                              \!\!\!\! per
                                              thread \end{tabular}}
                                  \\ \hline  
                                  Step~\ref{step:matrix_vector_product}:
                                  & $O(NT\bar{n}^{2})$           &
                                                                   $O(T\bar{n}^{2})$
                                  & $O(T\bar{n})$ %$O(T\bar{n})$
                                  \\ \hline
                                  Step~\ref{step:alpha_update}:
                                  & $O(NT\bar{n})$               &
                                                                   $O(T\bar{n})$ %+ O(N)$
                                  & $O(1)$                 \\ \hline
                                  Step~\ref{step:variable_update}:        & $O(NT\bar{n})$               & $O(T\bar{n})$               & 0                      \\ \hline
                                  Step~\ref{step:residual_update}:        & $O(NT\bar{n})$               & $O(T\bar{n})$               & 0                      \\ \hline
                                  Step~\ref{step:error_update}:
                                  & $O(NT\bar{n})$               &
                                                                   $O(T\bar{n})$ %+ O(N)$
                                  & $O(1)$                 \\ \hline
                                  Step~\ref{step:preconditioning}:
                                  & $O(LT\bar{n}^{3})$          &
                                                                   $O(LT\bar{n}^{3})$  
                                  & $O(LT\bar{n})$ %$O(T\bar{n})$
                                  \\ \hline
                                  Step~\ref{step:beta_update}:
                                  & $O(NT\bar{n})$               &
                                                                   $O(T\bar{n})$
                                                                   %  + O(N)$ 
                                  & $O(1)$                 \\ \hline
                                  Step~\ref{step:direction_update}:
                                  & $O(NT\bar{n})$               &
                                                                   $O(T\bar{n})$
                                  & 0
                                  \\ \hline
				\end{tabular}\\
				\caption{Complexity analysis of proposed PCG
                                  Algorithm~\ref{alg:pcg}: $\bar{n} =
                                  \max_j(n_j)$, where $n_j$ is the size
                                    of $x_{j,t}$; and
                                    $L$ is the fixed number
                                    of Jacobi iterations.}
				\label{tab:PCG_complexity_analysis}
			\end{table}
			\begin{rem}
                          The per PCG iteration computational
                          complexity is dominated by
                          step~\ref{step:preconditioning}, i.e.,
                          $O(LNT\bar{n}^3)$. With
                          the number $L$ of block Jacobi
                          preconditioning iterations fixed, and fixed bound
                          $\bar{n}$ on the size of sub-system states, the
                          overall computational complexity of PCG steps is
                          $O(NT)$.
			\end{rem}
			\begin{rem}
                          Note that steps~\ref{step:alpha_update},
                          \ref{step:error_update} and
                          \ref{step:beta_update} require sequential
                          computations, to accumulate in forming
                          dot-products and to test the stopping
                          condition. For the $\lceil N/2
                          \rceil$ parallel thread implementation,
                          these can be carried out using a
                          backward-forward sweep with path-graph data
                          exchange. Further, the parallel
                          implementation of
                          steps~\ref{step:matrix_vector_product} and
                          \ref{step:preconditioning} requires the
                          exchange of vectors of size less than
                          $T\bar{n}$, between the neighbouring threads
                          on this path-graph, since the partition of
                          $\Psi$ is block tri-diagonal. As such, the
                          overall inter-thread scalar data exchange
                          overhead is
                          $O(LNT\bar{n})$ per PCG iteration.
			\end{rem}
		%%% End of Section
	\section{Numerical Results}\label{sec:numerical_results}
        Numerical experiments are performed for an optimal control
        problem involving a one-dimensional mass-spring-damper chain
        of varying length of $N >
        0$ masses, taken from~\cite{Guo2019structured}. Each
        sub-system $j
        \in\mathcal{N}$ has dynamics of the form
        \eqref{eq:discrete_time_dynamics_cascaded_system} with $n_j =
        2$, $m_j = 1$, and $\nu_j =
        4$. The corresponding cost has $Q_{j,t} =
        \mathrm{diag}(1,0)$ and $R_{j,t} = 1$ for $t \in
        \mathcal{T}$. The model parameters such as mass, spring
        constant, damping coefficient are selected randomly between
        $0.8$ to
        $1.5$ to generate heterogeneous sub-systems. The experiments
        are performed by taking $N =
        T$ and varying this value from $10$ to
        $1000$. The number of scalar variables in the largest problem
        is in the order of
        $10^7$, and there are a similar number of constraints. The
        linear system of equations at each Newton-step is solved in
        the following ways:
		\begin{itemize}
                \item Algorithm~\ref{alg:pcg} to solve
                  \eqref{eq:block_tridiagonal_PD_reduced_kkt_equation}
                  with $L = 2$;
                  %, i.e., two iterations of block Jacobi
                  %method are performed for preconditioning at
                  %step~\ref{step:preconditioning};
                \item The block Jacobi method to solve
                  \eqref{eq:block_tridiagonal_PD_reduced_kkt_equation}
                  via iterations of the form
                  \eqref{eq:jacobi_iterations};
                \item The direct method~\cite{Cantoni2017structured}, via
                  backward-forward recursions (BFR) to effectively solve
                  \eqref{eq:reduced_kkt_equation} by LDL factorization;
                \item Solution of \eqref{eq:reduced_kkt_equation}
                  via MATLAB's backslash.
                  %with sparse arguments.
		\end{itemize}
		In order to gauge the overall computational complexity a
                single thread implementation is used for all
                methods.
                %That is, the possibility of exploiting
                %parallelism in Algorithm~\ref{alg:pcg} is not
                %considered.
                The duality-gap based stopping criterion for the interior
                point method is set to
                $\epsilon_{\mathrm{IPM}} = 10^{-6}$. The stopping
                criterion for the infinity norm of the residuals
                in Algorithm~\ref{alg:pcg}, and in the pure
                block Jacobi iterations based implementation, is set to
                $\epsilon = 10^{-9}$. For all experiments, IPM
                converged to specified tolerance within $15$ to $20$
                Newton steps.
		
\begin{figure*}[htp]
                  \centering
                  \begin{minipage}[b]{0.32\textwidth}
                  	\centering
                    \includegraphics[trim=5 3 32 21, clip, width=0.99\columnwidth]{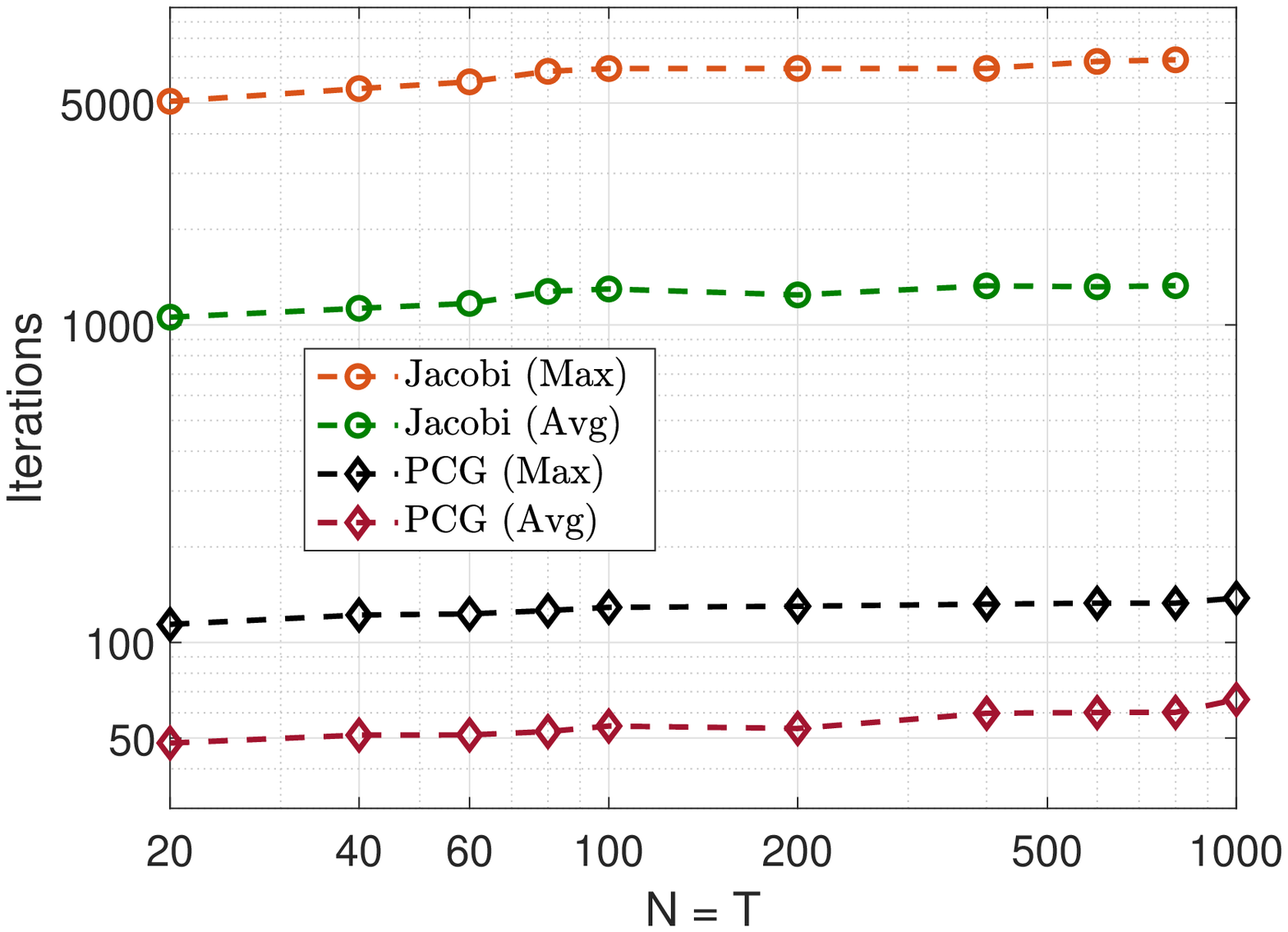}
                    \caption{Max/average iterations per Newton step}
                    \label{fig:N_T_20-1000_Tol_IPM_1e-06_Tol_PCG_1e-09_PCG_TAC_Ver4_BFRSolve_VPC_PCG_Iter}
                  \end{minipage}\hfill
                  \begin{minipage}[b]{0.32\textwidth}
                  	\centering
                          \includegraphics[trim=5 3 32 21, clip, width=0.99\columnwidth]{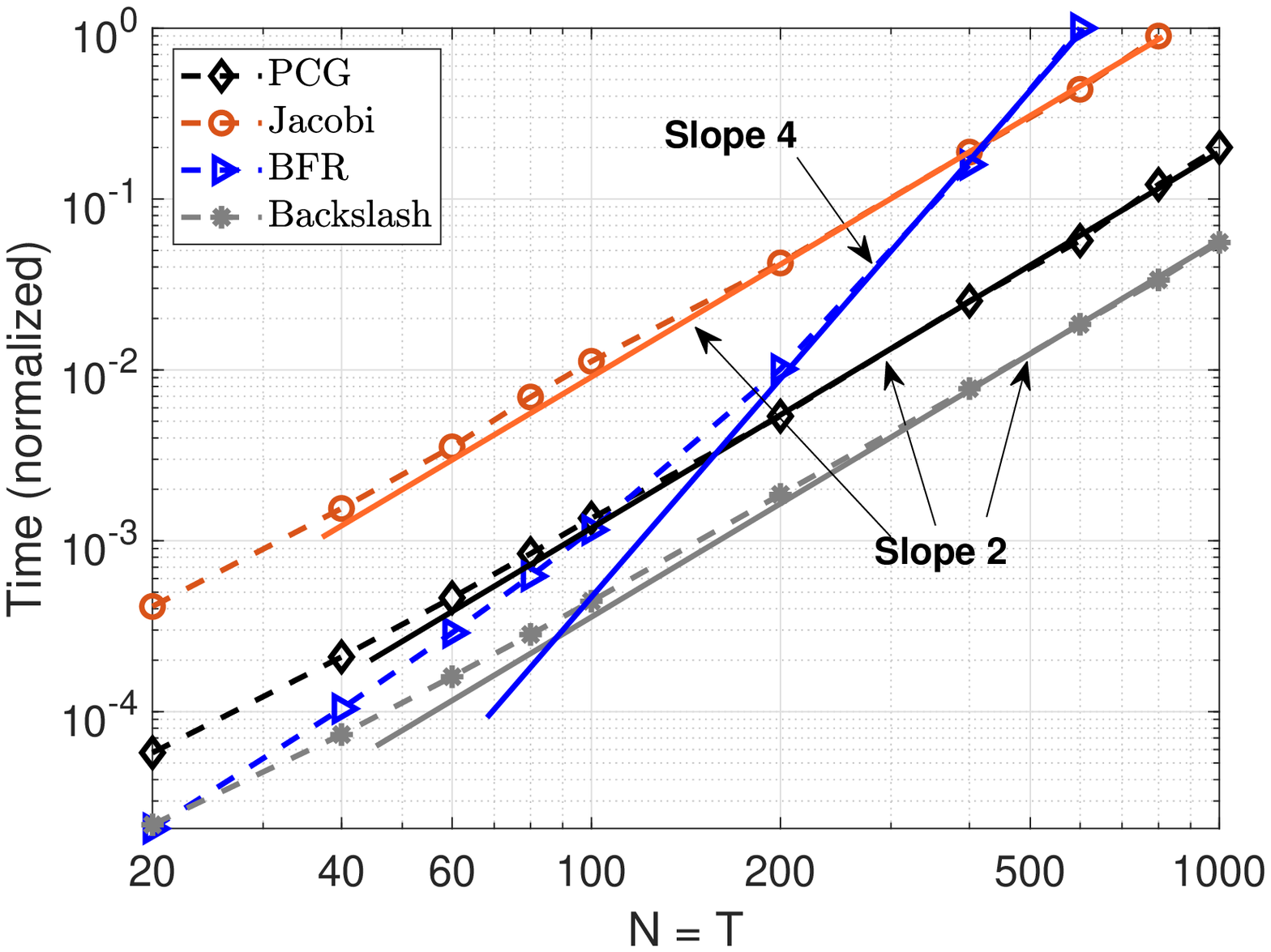}
                          \caption{Normalized average
                            time per Newton step} \label{fig:N_T_20-1000_Tol_IPM_1e-06_Tol_PCG_1e-09_PCG_TAC_Ver4_BFRSolve_VPC_Time_normalized}
                  \end{minipage}\hfill
                  \begin{minipage}[b]{0.32\textwidth}
                  	\centering
                    \includegraphics[trim=5 3 32 21, clip, width=0.99\columnwidth]{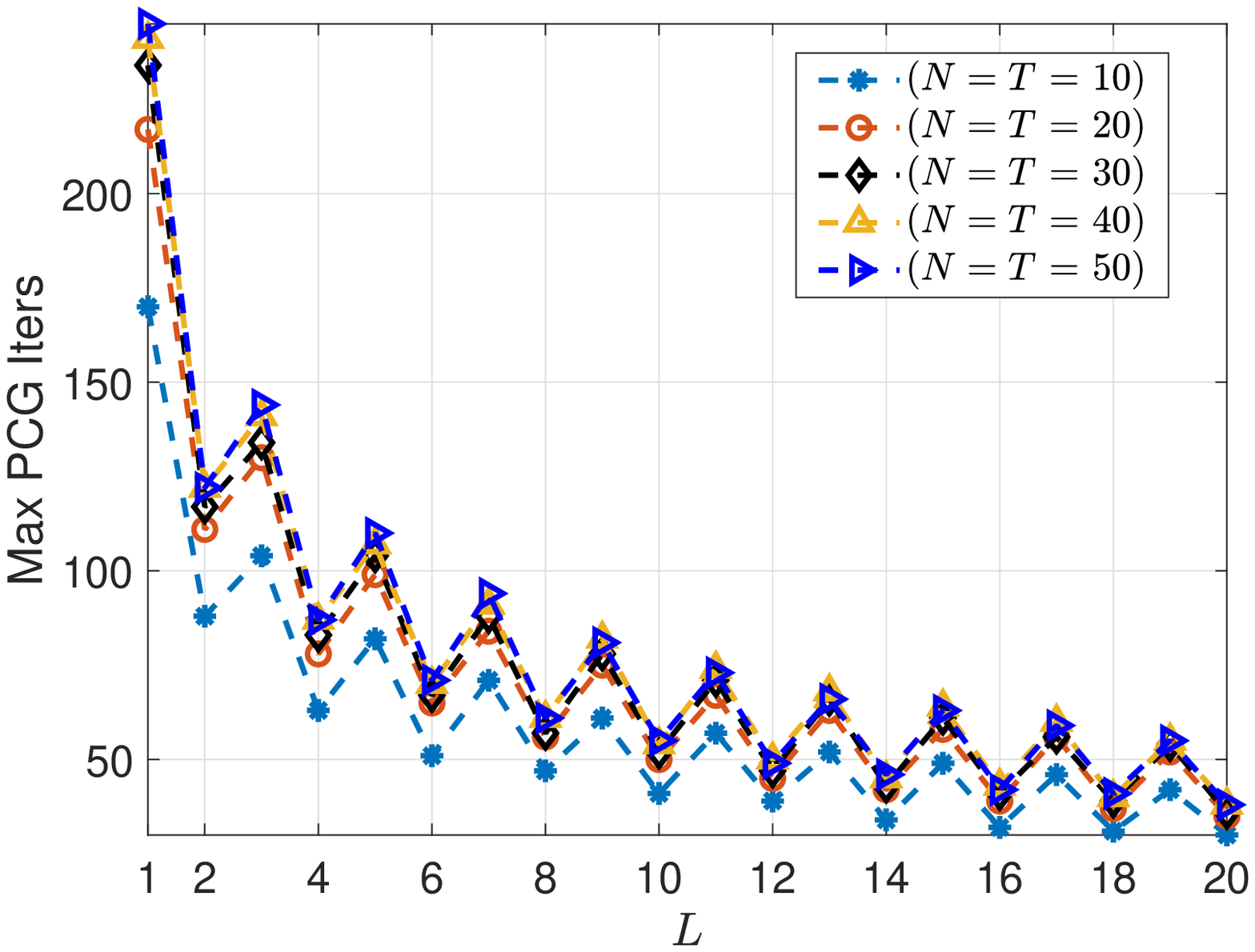}
                    \caption{Maximum PCG iterations vs. $L$}
                    \label{fig:N_T_10-50_Tol_IPM_1e-06_Tol_PCG_1e-09_msteps_1-20_PCG_TAC_Max_PCG_Iter}
                  \end{minipage}
                \end{figure*}

		Fig.~\ref{fig:N_T_20-1000_Tol_IPM_1e-06_Tol_PCG_1e-09_PCG_TAC_Ver4_BFRSolve_VPC_PCG_Iter}
                shows the maximum/average number of iterations for the
                pure block Jacobi method, and the PCG method with $L=2$, taken across
                IPM iterations. The pure block Jacobi method consistently involves a
                large number of iterations, in the order of
                thousands. By contrast, the proposed PCG method
                consistently requires far fewer iterations, in the
                order of hundreds. This demonstrates effectiveness of proposed
                approach to preconditioning.
                
		Fig.~\ref{fig:N_T_20-1000_Tol_IPM_1e-06_Tol_PCG_1e-09_PCG_TAC_Ver4_BFRSolve_VPC_Time_normalized}
                shows the normalized average processor time for a
                single thread implementation as proxy for the per-IPM
                iteration computational complexity. Along the line
                $N=T$, the average time is $O(N^2)$ for the PCG
                method, compared to $O(N^4)$ for the direct method
                \cite{Cantoni2017structured}. While the Jacobi method
                is also $O(N^2)$, the time is an order of magnitude
                greater than the PCG method. The average time for
                MATLAB’s backslash, based on MA-57
                \cite{Duff2004ma57}, is provided as a base line. Note,
                that backslash is able to permute matrices in ways
                that does not respect the spatio-temporal structure of
                problem \eqref{eq:LQ_optimal_control_problem}, which
                is by contrast preserved in the proposed PCG method.
		
		Finally, the effect of increasing $L$
                is shown in
                Fig.~\ref{fig:N_T_10-50_Tol_IPM_1e-06_Tol_PCG_1e-09_msteps_1-20_PCG_TAC_Max_PCG_Iter},
                as the value of $N = T$ is varied from $10$ to
                $50$. It can be seen that the maximum number of PCG
                iterations decreases as $L$ is increased, with
                considerable decrease as $L$ is increased from $1$ to
                $2$ for this example.
	%%%%%% end of section
	\section{Conclusions}\label{sec:conclulsions}
        A decomposable PCG method is proposed for computing
        second-order search directions for optimal control problems
        with path-graph network structure.
        The proposed algorithm exhibits per PCG iteration
        computational complexity that scales linearly with the number
        of sub-systems $N$ and the length of time horizon $T$. The
        computations at each iteration can be distributed across
        parallel processing agents in a network with path-graph
        structured information exchange. Future work includes
        extending the results for
        tree networks, where
        structure is manifest in three dimensions.
	%%%%%% end of section
	\bibliographystyle{IEEEtran}
%	\bibliography{library}

\end{document}